\documentclass[article, shortnames]{jss}
\usepackage{amsmath}
\usepackage{graphicx}

\author{Stephen Reid\\Stanford University \And 
        Rob Tibshirani\\Stanford University}
\title{Regularisation Paths for Conditional Logistic Regression: the \pkg{clogitL1} package}

\Plainauthor{Stephen Reid, Rob Tibshirani} 
\Plaintitle{Regularisation Paths for Conditional Logistic Regression: the \pkg{clogitL1} package} 
\Shorttitle{\pkg{clogitL1}: Regularisation Paths for Conditional Logistic Regression} 

\Abstract{
  We apply the cyclic coordinate descent algorithm of \citet{FHT2010} to the fitting of a conditional logistic regression model with lasso ($\ell_1$) and elastic net penalties. The sequential strong rules of \citet{TBFHSTT2012} are also used in the algorithm and it is shown that these offer a considerable speed up over the standard coordinate descent algorithm with warm starts.

Once implemented, the algorithm is used in simulation studies to compare the variable selection and prediction performance of the conditional logistic regression model against that of its unconditional (standard) counterpart. We find that the conditional model performs admirably on datasets drawn from a suitable conditional distribution, outperforming its unconditional counterpart at variable selection. The conditional model is also fit to a small real world dataset, demonstrating how we obtain regularisation paths for the parameters of the model and how we apply cross validation for this method where natural unconditional prediction rules are hard to come by. 
}
\Plainkeywords{conditional logistic regression, elastic net, lasso, cyclic coordinate descent, sequential strong rules} 

\Address{
  Stephen Reid\\
  Department of Statistics\\
  Stanford University\\
  390 Serra Mall, Stanford, CA, USA\\
  E-mail: \email{sreid@stanford.edu}\\

  Rob Tibshirani\\
  Department of Statistics\\
  Stanford University\\
  390 Serra Mall, Stanford, CA, USA\\
  E-mail: \email{tibs@stanford.edu}\\
  URL: \url{http://www-stat.stanford.edu/~tibs}  
}

\begin{document}

\section{Introduction}
Suppose we have $K$ independent strata, each with $n_k$ independent observations, $(Y_{ki}, X_{ki})$, $k = 1, 2, ..., K; i = 1, 2, ..., n_k$, where each $Y_{ki}$ is a dichotomous random variable taking on values $0$ or $1$ and each $X_{ki}$ is a $p$-vector of fixed regressors or exposure variables. The quantities are related via the logistic regression model:

\begin{equation*}\label{eq1.1}\tag {1.1}
 \mathrm{logit} P(Y_{ki} = 1|X_{ki}) \doteq \log\left( \frac{P(Y_{ki} = 1|X_{ki})}{1 - P(Y_{ki} = 1|X_{ki})}\right) = \beta_{0k} + \beta^\top X_{ki}
\end{equation*}

Note that each observation has two indices: the first identifying its stratum; the second, its number within the stratum. The stratum effect is incoporated by giving each stratum its own intercept $\beta_{0k}$ (with $\beta$ the same for all strata). The logistic regression model has been studied extensively and its properties are well established and widely known.

Furthermore, consider a retrospective (case-control) study design where the number of cases $(Y_{ki} = 1)$ and controls $(Y_{ki} = 0)$ in each stratum $k$ are fixed beforehand at $m_k$ and $n_k-m_k$ respectively. We may assume, without loss of generality, that the observations at indices $i = 1, 2, ..., m_k$ in stratum $k$ are the cases and those at indices $i = m_k + 1, m_k + 2, ..., n_k$ are controls. These types of studies are more feasible in practice when prospective studies are too costly, time consuming or simply unethical. Examples abound in epidemiology, economics and the actuarial sciences.

Now the core assumption of model Equation~\ref{eq1.1} - $Y$ random; $X$ fixed - is exactly reversed and the likelihood that stems from that equation is no longer valid for the probability mechanism generating the data. In the conditional logistic likelihood literature, this problem is dealt with by treating the intercepts $\beta_{0k}$ as nuisance parameters and making a conditional argument to derive a new quantity, called the ``conditional logistic likelihood'':

\begin{align*}\label{eq1.2}
   L(\beta) &= \prod_{k = 1}^KL_k(\beta) \\
   &= \prod_{k = 1}^K\frac{\exp(\beta^\top\sum_{i=1}^{m_k}X_{ki})}{\sum_{u \in S_{m_k,n_k}}\exp(\beta^\top\sum_{i \in u}X_{ki})}\tag{1.2}
\end{align*}

See \citet{BD80} for further details. The set $S_{m, n}$ is defined in Section 2.

We maximise Equation~\ref{eq1.2} to obtain estimates for $\beta$. The advantage of using the conditional likelihood is twofold: 

\begin{enumerate}
  \item All nuisance parameters are eliminated, possibly reducing the standard error and bias of the estimates of $\beta$.
  \item This conditional probability (and hence the likelihood) remains valid regardless of study type (prospective or retrospective).
\end {enumerate}

This method is used in a variety of application areas to study the effect of exposure variables on some dichotomous event of interest. An example is in epidemiology where we have data on $K$ patients (our strata) and tissue characteristics for different tissue samples from each patient. For each patient, some tissue samples are cancerous, while others are healthy. The goal would be to find those tissue characteristics most related to the development of cancer.

It seems natural to have some principled, automatic method for selecting the relevant exposure variables. Here we propose a penalised conditional logistic regression model where we minimise:

\begin{equation}\tag{1.3}\label{eq1.3}
  -\log L(\beta) + \lambda P_\alpha(\beta)
\end{equation}
where

\begin{equation}\tag{1.4}\label{eq1.4}
\lambda P_\alpha(\beta) = \lambda\left(\alpha \sum_{j = 1}^p |\beta_j| + \frac{1}{2}(1-\alpha)\sum_{j = 1}^p\beta_j^2\right)
\end{equation}
is the elastic net penalty proposed by \citet{ZH2005}.

In this paper, we apply the cyclic coordinate descent algorithm of \citet{FHT2010} and \citet{WL2007a} to obtain a path of penalised solutions. Having a path of solutions facilitates cross validation when determining the optimal $\lambda$ value. A recursive formula proposed by \citet{Gail81} is used to compute the likelihood and its derivatives exactly. The cyclic coordinate descent algorithm is stable and reasonably efficient.

Section 2 discusses the model rationale and algorithm implementation: how we compute the normalising constant in the denominator of Equation~\ref{eq1.2} and how we apply cyclic coordinate descent to obtain the solution path. Section 3 shows how sequential strong rules drastically improve the efficiency with which we can compute the solutions. Section 4 looks at the time taken to arrive at solutions for datasets of different sizes. A comparison is made between the conditional and unconditional models in Section 5, looking at the difference in variable selection performance and predictive power. Finally, Section 6 briefly addresses the implementation of cross validation for a method that does not immediately present us with a convenient way of making predictions.

We provide a publicly available \proglang{R} (\citet{Rlang}) implementation in the package \pkg{clogitL1}, available from the Comprehensive \proglang{R} Archive Network (CRAN) at \url{http://CRAN.R-project.org/package=clogitL1}.

\section{Rationale and algorithm}
\citet{BD80} suggest three approaches to adapt the standard logistic model to the case-control design. The first involves some assumptions on the effect of regressors $X_{ki}$ on the probability of being selected to the sample (some of which are unlikely to be bourne out in practice). It can be shown that the standard logistic model (with slightly different values for the intercepts) can be applied with impunity in this case.

Their second suggestion uses Bayes' rule to justify the use of the standard logistic model. However, one still has to contend with many nuisance parameters (intercepts and marginal distributions of $X$ and $Y$). 

The approach adopted here is to use a suitable conditional probability to eliminate the intercepts and finesse the need to estimate fully general marginal distributions. Attention focuses, in each stratum $k$, on the conditional probability:

\begin{equation}\label{eq2.1}\tag{2.1}
  P(Y_{k1} = Y_{k2} = ... = Y_{km_k} = 1 | \sum_{i=1}^{n_k}Y_{ki} = m_k) = 
                \frac{\exp(\beta^\top\sum_{i=1}^{m_k}X_{ki})}{\sum_{u \in S_{m_k,n_k}}\exp(\beta^\top\sum_{i \in u}X_{ki})}
\end{equation}
where $S_{m, n}$ is the collection of ${n \choose m}$ sets $\{i_1, i_2, ..., i_m\}$ where $1 \leq i_i < i_2 < ... < i_m \leq n$. This is the conditional probability that the case indices are as observed given that exactly $m_k$ cases were observed.

Assuming that observations are independent across strata, the conditional likelihood is as in Equation~\ref{eq1.2}.

We wish to find $\beta$ that minimises a penalised version of the (negative) log conditional likelihood: 

\begin{equation*}\label{logLik}
  -l(\beta) = -\beta^\top\sum_{i=1}^{m_k}X_{ki} + \log\left(\sum_{u \in S_{m_k,n_k}}\exp(\beta^\top\sum_{i \in u}X_{ki})\right)
\end{equation*}

The Lagrangian formulation of the problem becomes:

\begin{equation}\label{eq2.2}\tag{2.2}
 \hat{\beta} = \mathrm{argmin}_\beta\left[-\beta^\top\sum_{i=1}^{m_k}X_{ki} + \log\left(\sum_{u \in S_{m_k,n_k}}\exp(\beta^\top\sum_{i \in u}X_{ki})\right) + \lambda P_\alpha(\beta)\right]
\end{equation}
where $P_\alpha(\beta)$ is as in Equation~\ref{eq1.4}.

The $\alpha$ parameter trades off between the $\ell_1$ and $\ell_2$-penalties, the former encouraging sparsity (setting individual $\beta_j$ estimates to $0$), while the latter merely shrinks estimates toward $0$. The default is $\alpha = 1$, which corresponds to a pure $\ell_1$-penalty. As we decrease $\alpha$ toward $0$, we retain the variable selecting behaviour of an $\ell_1$-penalty while improving stability of $\beta$ estimates.

The parameter $\lambda$ controls the extent of regularisation. As $\lambda \to \infty$, $\hat{\beta}$ is eventually set to $0$, while the maximum conditional likelihood estimate is obtained when $\lambda = 0$. As $\lambda$ increases from $0$, we obtain solutions with progressively fewer non-zero parameter estimates.

\subsection{A small example}

Before delving into the technical details, let us consider a small real world example. The dataset under consideration is from \citet{BD80}. It comprises of $315$ observations divided into $63$ strata of size $5$. Each stratum contains one case (a patient with endometrical cancer) and $4$ controls (non-cancer patients), matched by date of birth, marital status and residence. Since observations come in matched sets, conditional logistic likelihood maximisation is ideally suited here.

Predictor variables are rife with missing values and we selected only $5$ out of a possible $11$ predictors for this small study. Predictors included in this example are:

\begin {itemize}
  \item \emph{gall}: Indicator of gall bladder disease.
  \item \emph{hyp}: Indicator of hypertension.
  \item \emph{est}: Indictor of use of oestrogen drugs.
  \item \emph{non}: Indicator of use of non oestrogen drugs.
  \item \emph{age}: Age in years.
\end{itemize}

The \emph{age} variable was centered and standardised to ensure that the scale of this variable is comparable to that of the indicator variables. The scale of predictors affect their $\ell_1$-penalised parameter estimates.

\begin{figure}[htb]
  \centering
  \includegraphics [height = 100mm, width = 100mm] {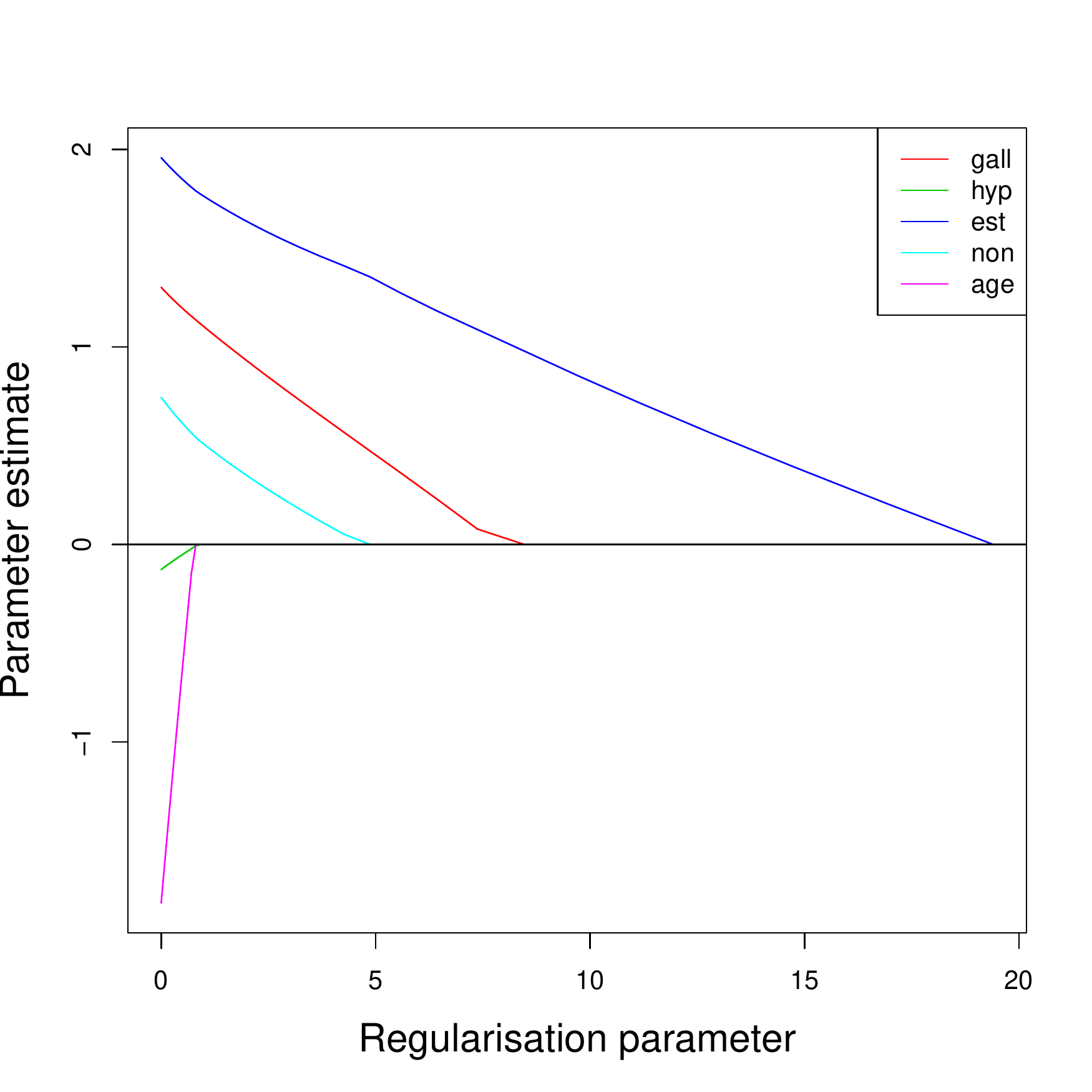}
  \caption {\emph{Parameter profile for endometrical cancer dataset. $\beta$ estimates are plotted against the regularisation parameter $\lambda$.}}
  \label{fig1}
\end{figure} 

Figure~\ref{fig1} plots the profile of parameter estimates against the value of the penalty parameter $\lambda$ (we set $\alpha = 1$). We start at $\lambda$ for which all parameter estimates are $0$ (far right), decreasing $\lambda$ and recomputing estimates until we reach the unconstrained maximum conditional likelihood estimates ($\lambda = 0$; far left).

Gall bladder disease and the taking of oestrogen and non-oestrogen drugs all seem to have a positive effect on the probability of developing cancer. The oestrogen drug indicator is the first predictor to have non-zero parameter estimate and seems to have a significant effect on cancer development. Age and hypertension seem to have modest negative effects on cancer development, but both of these predictors enter late and are probably not significant. 

Now let us discuss the technical details behind the computation of a parameter profile as in Figure~\ref{fig1}.

\subsection{Penalised least squares}
We employ a strategy very similar to the standard Newton-Raphson algorithm for solving Equation~\ref{eq2.2}. At each iteration, rather than solving a standard least squares problem, we solve a penalised one.

The quadratic approximation of the log conditional likelihood centered at a point $\tilde{\beta}$ is

\begin{align*}
  l(\beta) &= l(\tilde{\beta}) + (\beta - \tilde{\beta})^\top \dot{l}(\tilde{\beta}) + \frac{1}{2}(\beta - \tilde{\beta})^\top \ddot{l}(\tilde{\beta})(\beta - \tilde{\beta}) 
\\
  &= l(\tilde{\beta}) + (\beta - \tilde{\beta})^\top s - \frac{1}{2}(\beta - \tilde{\beta})^\top H(\beta - \tilde{\beta})
\end{align*}
where $s = \dot{l}(\tilde{\beta})$ is the $p$-vector of first derivatives (score vector) and $H = - \ddot{l}(\tilde{\beta})$ is the (negative) Hessian. The form and computation of these quantities are discussed later.

The algorithm becomes:
\begin{enumerate}
  \item Initialise $\tilde{\beta}$.
  \item Compute $s$ and $H$.
  \item Find $\hat{\beta}$ minimising $-(\beta - \tilde{\beta})^\top s + \frac{1}{2}(\beta - \tilde{\beta})^\top H(\beta - \tilde{\beta}) +  \lambda P_\alpha(\beta)$.
  \item Set $\tilde{\beta} = \hat{\beta}$.
  \item Repeat steps 2 through 4 until convergence.
\end{enumerate}

In principle, this algorithm could be implemented with the solver of one's choice. The criterion in step 3 is convex in the optimisation variable and any effective convex optimisation solver (with the necessary tweaks required of these) would converge to the global optimum of the objective. We, however, implement a different algorithm. Section 2.3 describes the minimisation in step 3, which is done via cyclic coordinate descent.

\subsection{Cyclic coordinate descent}
The minimisation in step 3 above is done one coordinate at a time, rather than in a batch as in the standard Newton-Raphson update step. Letting $M(\beta)$ denote the objective in step 3, we obtain the derivative:

\[
\frac{\partial M}{\partial \beta_j} = -s_j + (\beta_j - \tilde{\beta}_j)H_{jj} + \sum_{l \not= j}(\beta_l - \tilde{\beta}_l)H_{lj} + \lambda\alpha \cdot \mathrm{sign}(\beta_j)+ \lambda (1-\alpha) \cdot \beta_j
\]
Setting each of these equations to $0$, we obtain the following update:

\begin{equation}\label{eq2.3}\tag{2.3}
  \hat{\beta_j} = \frac{S\left(\tilde{\beta_j}H_{jj} + s_j + \sum_{l \not= j}(\tilde{\beta_l} - \hat{\beta_l})H_{lj}, \lambda\alpha \right)}{H_{jj} + \lambda(1-\alpha)}
\end{equation}
where

\[
 S(x, t) = \mathrm{sign}(x) \cdot (x - t)_+
\]
is the soft thresholding operator.

These are cycled over $t = 1, 2..., p, 1, 2, ..., p, ...$ until convergence, each time updating the value of $\hat{\beta}$ for use in subsequent updating equations. Note that $\tilde{\beta}$, $s$ and $H$ remain fixed throughout this cyclic coordinate descent epoch.

Convergence of this algorithm is discussed in \citet{FHT2010}. Their sentiments will largely be reiterated in Section 2.6, once we have discussed the pathwise algorithm.

\subsection{Computing the likelihood and its derivatives}
Our cyclic coordinate descent updates require computation of the entire score vector, the diagonal of the Hessian and off diagonal Hessian entries for those $j$ such that $\tilde{\beta_j}$ and $\hat{\beta_j}$ become unequal during the descent phase.

Taking first and second derivatives of the likelihood in Equation~\ref{eq1.2}, we obtain the following expressions for the score vector and (negative) Hessian:

\[
s = \sum_{k=1}^K\left(\sum_{i=1}^m X_{ki} - \mu_k\right)
\]

\[
H = \sum_{k=1}^K\sum_{u \in S_{m,n}}p_{ku}\left(\sum_{i \in u}X_{ki} - \mu_k\right)\left(\sum_{i \in u}X_{ki} - \mu_k\right)^\top
\]
where 

\[
\mu_k =  \sum_{u \in S_{m,n}}p_{ku}\sum_{i \in u}X_{ki}
\]
and
\[
 p_{ku} = \frac{\exp(\beta^\top \sum_{i \in u} X_{ki})}{\sum_{v \in S_{m, n}}\exp(\beta^\top\sum_{i \in v}X_{ki})}
\]

Brute force computation of $s$ and $H$ requires enumeration of at least $n_k \choose m_k$ terms in each stratum, leading to an $O({n_k \choose m_k})$ computation time in each stratum $k$ for every entry of $s$ and $H$ (of which there are $O(p^2)$). Computation increases rather rapidly as $n_k$ and $m_k$ increase. For example, ${20 \choose 10} = 184756$ and ${40 \choose 20} = 1.378 \times 10^{11}$. It does not seem implausible that strata should contain as many as $40$ (or more) observations. A particular medical application could allow for measuring many tissue samples per patient. 

Even though \citet{BD80} suggest that $\beta$ estimates and their standard errors from the conditional likelihood are almost certain to be numerically close to those from the standard unconditional likelihood for large $n_k$ and $m_k$, the arguments are asymptotic. What is large enough to ensure this numerical similarity is often to be judged for each application separately and then one needs a method to compute the conditional likelihood estimates. A more amenable method for computing these quantities is required.

\subsection{The normalising constant and its derivatives}
The major computation bottleneck occurs because the normalising constant in each stratum comprises of $n_k \choose m_k$ terms. These are carried over into the first and second derivatives, because the derivative operator commutes with finite sums.

Consider the normalising constant for a single stratum $k$, omitting subscripts for ease of exposition:

\begin{equation}\label{eq2.4}\tag{2.4}
 B_k(m, n) = \sum_{u \in S_{m,n}}\exp(\beta^\top\sum_{i \in u} X_{ki})
\end{equation}

\citet{Gail81} propose a recursive formula for the computation of Equation~\ref{eq2.4}:

\begin{equation}\label{eq2.5}\tag{2.5}
B_k(m, n) = B_k(m, n-1) + e^{\beta^\top X_{kn}}B_k(m-1, n-1)
\end{equation}
with base case conditions:  $B_k(0, n) = 1 \forall n$ and $B_k(m, n) = 0$ $m > n$. 

The intuition is that, to compute the normalising constant for a stratum with $n$ observations of which $m$ are cases, we can fix attention on the last observation (numbered $n$). The overall sum can be split into one which does not contain a term relating to observation $n$ and one which does. The former sum is simply the normalising constant of a stratum with $n-1$ observations from which we still require $m$ cases (see first term of Equation~\ref{eq2.4}). The second is the normalising constant for a stratum with $n-1$ observations from which we require only $m-1$ cases, because we have already kept one observation aside as a case. We must multiply this second normalising constant by the contribution of this kept out observation, hence the second term.

Using the recursive formula allows us to compute the normalising constant in $O(m(n-m))$ time, suggesting that the entire likelihood can be computed in $O(\sum_{k=1}^Km_k(n_k - m_k))$ time.

Taking first and second derivatives of Equation~\ref{eq2.5} on both sides yields recursive expressions for the first and second derivatives of the normalising constant:

\[
\dot{B}_k(m,n) = \dot{B}_k(m, n-1) + e^{\beta^\top X_{kn}}\dot{B}_k(m-1,n-1) + X_{kn}e^{\beta^\top X_{kn}}B_k(m-1, n-1)
\]
and 
\begin{align*}
\ddot{B}_k(m, n) = \ddot{B}_k(m, n-1) +  e^{\beta^\top X_{kn}} \ddot{B}_k(m-1,n-1) + X_{kn}e^{\beta^\top X_{kn}} \dot{B}_k(m-1,n-1)^\top
\\
+ e^{\beta^\top X_{kn}} \dot{B}_k(m-1,n-1)X_{kn}^\top + X_{kn}X_{kn}^\top e^{\beta^\top X_{kn}}B_k(m-1, n-1)
\end{align*}
The score vector and Hessian can be computed fairly directly once we have these quantities:
\[
s = \sum_{k=1}^K\left(\sum_{i=1}^m X_{ki} - \frac{\dot{B}_k(m_k, n_k)}{B_k(m_k, n_k)}\right)
\]

\[
H = \sum_{k=1}^K\left[\frac{\ddot{B}_k(m_k, n_k)}{B_k(m_k, n_k)^2} - \left(\frac{\dot{B}_k(m_k, n_k)}{B_k(m_k, n_k)}\right)\left(\frac{\dot{B}_k(m_k, n_k)}{B_k(m_k, n_k)}\right)^\top \right] 
\]

Each entry of $s$ and $H$ can now also be computed in $O(\sum_{k=1}^Km_k(n_k-m_k))$ time. Notice that the recursive calculations depend on those obtained during the computation of lower order derivatives. This has interesting implications for the implementation of the algorithm. 

Consider Equation Equation~\ref{eq2.3}. Without any prior knowledge of which parameter estimates are to become non-zero in the subsequent cyclic coordinate descent epoch, we are forced to compute $s$ and the diagonal elements of $H$. We can postpone the computation of the off diagonal elements until such time that $\hat{\beta_j}$ differs from $\tilde{\beta_j}$, at which point we compute all the off diagonal Hessian elements involving variable $j$. To implement this strategy, we would need to keep persistent copies of all the intermediate recursive computations obtained while computing the normalising constant and its first derivatives - a memory burden of $O(p\sum_{k = 1}^Km_k(n_k-m_k))$. For large $p$, $n_k$ and $m_k$ this could amount to an unacceptable use of memory. 

Alternatively, all entries of $s$ and $H$ could be computed before we begin our cyclic descent epoch. Computing these quantities at the same time would allow for intermediate computations to be discarded as soon as we have the final values. This, on the other hand, would lead to excessive computation time, with much time being spend wastefully, computing Hessian entries of many variables that are unlikely to become non-zero over the current cyclic descent epoch.

Ideally, we would like some heuristic that could help predict which elements of $\hat{\beta}$ are likely to become non-zero in the next epoch (hopefully far fewer than $p$ of them) and perform the whole batch of likelihood computations for only this comparatively small set of variables. We discuss this later.

\subsection{Pathwise algorithm}
We are interested in computing parameter estimates not at one particular value of $\lambda$, but rather a grid of, say, $100$ of them. To this end, we begin with the smallest $\lambda$ such that all parameter estimates are 0. From Equation~\ref{eq2.3}, this is easily seen to be:

\[
\lambda_{max} = \frac{\max_j|s_j|}{\alpha}
\]

We then find the optimal $\beta$ for each $\lambda$ along a grid of $100$ values spaced between $\lambda_{max}$ and $\lambda_{min} = \epsilon\lambda_{max}$ where $\epsilon = 0.00001$. Notice that we do not run down to the unregularised solution. For $p > \sum_{k=1}^Kn_k$ this solution is undefined anyway and for $p \le \sum_{k=1}^Kn_k$, solutions can be numerically unstable at $\lambda = 0$.

Each solution at the current $\lambda$ value is used as a warm start for the algorithm when computing the solution at the next $\lambda$ value. This lends stability to an algorithm that is not guaranteed to converge without step size optimisation (which we do not do here). No problems of such ilk have been encountered yet. Warm starting on previous solutions gives us a good chance of starting in the zone of quadratic convergence of the Newton-Raphson algorithm, where quadratic approximations to the objective are good and convergence is rapid and guaranteed.

\citet{FHT2010} suggest that $\lambda$ values should be spaced uniformly on the log scale, suggesting that successive $\lambda$ values are related via:

\[
 \lambda_{k+1} = \lambda_k\left(\frac{\lambda_{min}}{\lambda_{max}}\right)^{\frac{1}{100}}
\]

It was found that the nature of the grid has significant influence on the overall computation time. This is discussed further in the next section.

For the case where $\sum_{k = 1}^Kn_k < p$, one cannot estimate more than $\sum_{k = 1}^Kn_k$ parameters. We would like to terminate the algorithm early should we explain almost all of the variability in the observations. Following \citet{simon2011}, we use a deviance cutoff. 

The deviance of a model with parameter $\beta$ is defined to be:

\[
D(\beta) = 2(l_{saturated} - l(\beta))
\] 
where $l_{saturated}$ is the maximum log-likelihood value achieved when the exponents in the likelihood are allowed to vary freely and not just over linear combinations of $\beta$. Fairly simple arguments suggest that in this case $l_{saturated} = 0$.

The algorithm terminates when
\[
D_{null} - D(\beta_{current}) \geq 0.99D_{null}
\]
where $D_{null} = 2(0 - l(0)) = -2\sum_{k=1}^K\log{n_k \choose m_k}$.

\section{Strong rules}
It has been intimated in the previous section that the recursive structure of the computation of the likelihood and its derivatives encourages a search for a heuristic for reducing the number of score and Hessian entries to be computed in each iteration of our algorithm. Should we find a good heuristic, we could reduce the computation burden and simultaneously reduce the memory overhead by computing all related quantities in a batch. Obviously we would still like to obtain the exact penalised maximum conditional likelihood estimates after its application.

\citet{TBFHSTT2012} propose a heuristic for screening variables immediately after the most recent decrement of $\lambda$. Suppose we have just computed the solution at $\lambda_{k-1}$: $\hat{\beta}(\lambda_{k-1})$ and the associated score vector $s$. Their sequential strong rule suggests that, when solving for the current $\lambda_k$, we may ignore all variables such that:
\begin{equation}\label{eq3.1}\tag{3.1}
|s_j| \le \alpha(2\lambda_{k} - \lambda_{k-1})
\end{equation}
Steps 2 - 6 in Section 2.2 may then proceed with only those variables that pass this test. Those that do not pass the test are simply left at $0$.

The authors show that this rule is not safe and some parameter estimates that should be non-zero at the end of the iteration could be screened at the start of it. However, they also show that such errors are rare. Either way, one need only check that KKT conditions at the end of each iteration and rerun steps 2 - 6 should a violation be detected, using the current solution as a warm start. These KKT conditions are:

\[
 |s_j - \lambda_k(1-\alpha)\hat{\beta}_j(\lambda_k)| \le \lambda_k
\]
for $j = 1, 2,..., p$, where the score vector $s$ is computed at the proposed solution at $\lambda_k$. Should all variables satisfy the KKT conditions, we have found the exact penalised maximum conditional likelihood parameter estimates.

The sequential strong rule is shown to discard many variables over the course of the algorithm and reduces overall computation time considerably. For the current application, it has the additional benefit of precluding the need for persistent copies of intermediate recursive computations, reducing memory overhead. 

\subsection{Efficiency gains of sequential strong rules}
The sequential strong rule provides a significant speed up (often by a factor of 2 - 5 over the non-strong-rule case). Not only does it allow the computation of fewer Hessian entries per iteration, but it also allows a reordering of the score and Hessian computations, making the code more amenable to compiler optimisation.

\begin{figure}[htb]
  \centering
  \includegraphics[width = 100mm, height = 100mm]  {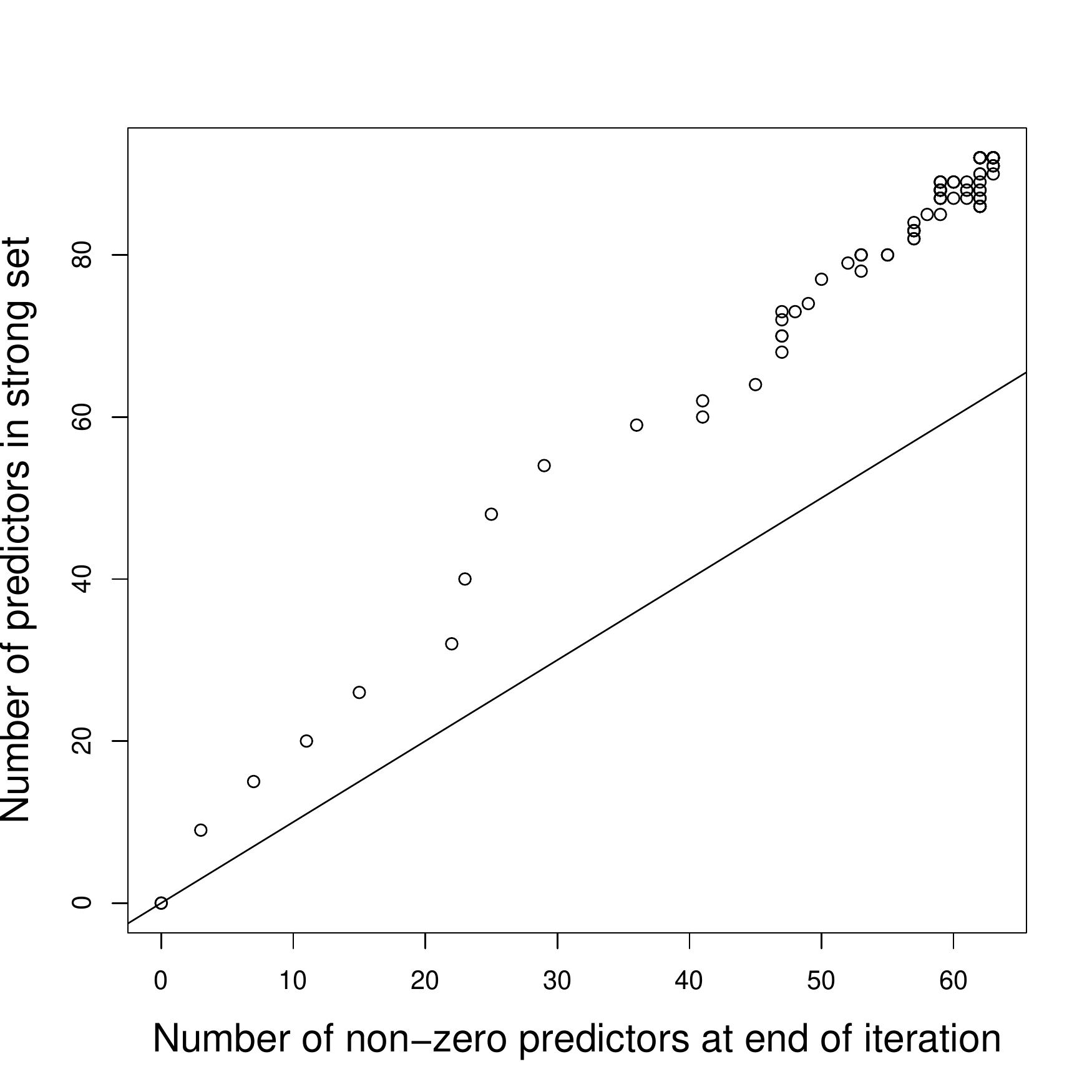}
  \caption{\emph{Screening performance of sequential strong rule. $K = 10$, $n = 10$, $m = 5$, $p = 200$.}}
  \label{fig2}
\end{figure}

Figure~\ref{fig2} shows the screening performance of the strong rule. The actual number of non-zero predictors comprising the solution at each $\lambda$ value is plotted on the horizontal axis. The number of predictors selected to the strong set (those on which we solely focus for that iteration) is plotted on the vertical axis. The unit slope line through the origin is plotted for reference. If the strong rule worked perfectly, points would lie along the reference line. We see that the strong rule is somewhat conservative, always selecting more variables for consideration than those that turn out to be active at the end of the iteration. 

A matter not addressed by \citet{TBFHSTT2012} is the cost of this conservatism. In their applications, score and Hessian computations are comparatively cheap. Here they are quite expensive and any savings on the number of these quantities computed is welcome. Perhaps then there is a way to choose the grid of $\lambda$ visited so that the sequential strong rule chooses smaller strong sets in each iteration. This is discussed in the next section.

\subsection{Choosing different grids}
We experimented with different ways of choosing the grid of $100$ $\lambda$ values. One obvious alternative to the uniform logarithmic grid is the uniform linear grid:

\[
\lambda_{k+1} = \lambda_k - \frac{\lambda_{max} -\lambda_{min}}{100}
\]

This grid has the unwelcome property that the strong rule threshold Equation~\ref{eq3.1} becomes negative for small $\lambda$, forcing all predictors into the strong set, considerably slowing down the algorithm in the final iterations.

Perhaps a compromise between the two grids could be tried, starting with linear jumps between successive $\lambda$ values and then transitioning to logarithmic jumps at later iterations.

A simulation study comparing the performance of the different grid choices was run. Fixing $\alpha = 1$ and the number of strata at $K = 10$, as well as the number of observations and cases per stratum at $n = 10$ and $m = 5$, respectively, regressors $X_{kij}$, $k = 1, 2, ..., K$, $i = 1, 2, ..., n$, $j = 1,2, ...,p$ were generated independently from a $N(0,1)$ distribution. The number of predictors was set to $p = 100, 200, 500, 1000, 2000$ in turn. $q = \lfloor0.25p\rfloor$ of the regression parameters were chosen randomly, assigned a sign uniformly at random and these comprised the non-zero predictors (with absolute value $|\beta_j| = 2$). For each stratum, $m$ cases were selected (without replacement) based on the logistic success probabilities implied by the generated regressors and coefficients.

$B = 3$ of these datasets were generated for each $p$ and the algorithm was put to work at computing a path of $\beta$ solutions over $4$ different grids of $100$ $\lambda$ values. All simulations were done on an Intel Dual Core 2.5GHz processor.

Table~\ref{tabDiffGrids} shows the average running times for each of 4 different grid types.

\begin{figure}[htb]
    \centering
    \includegraphics [height = 100mm, width = 100mm] {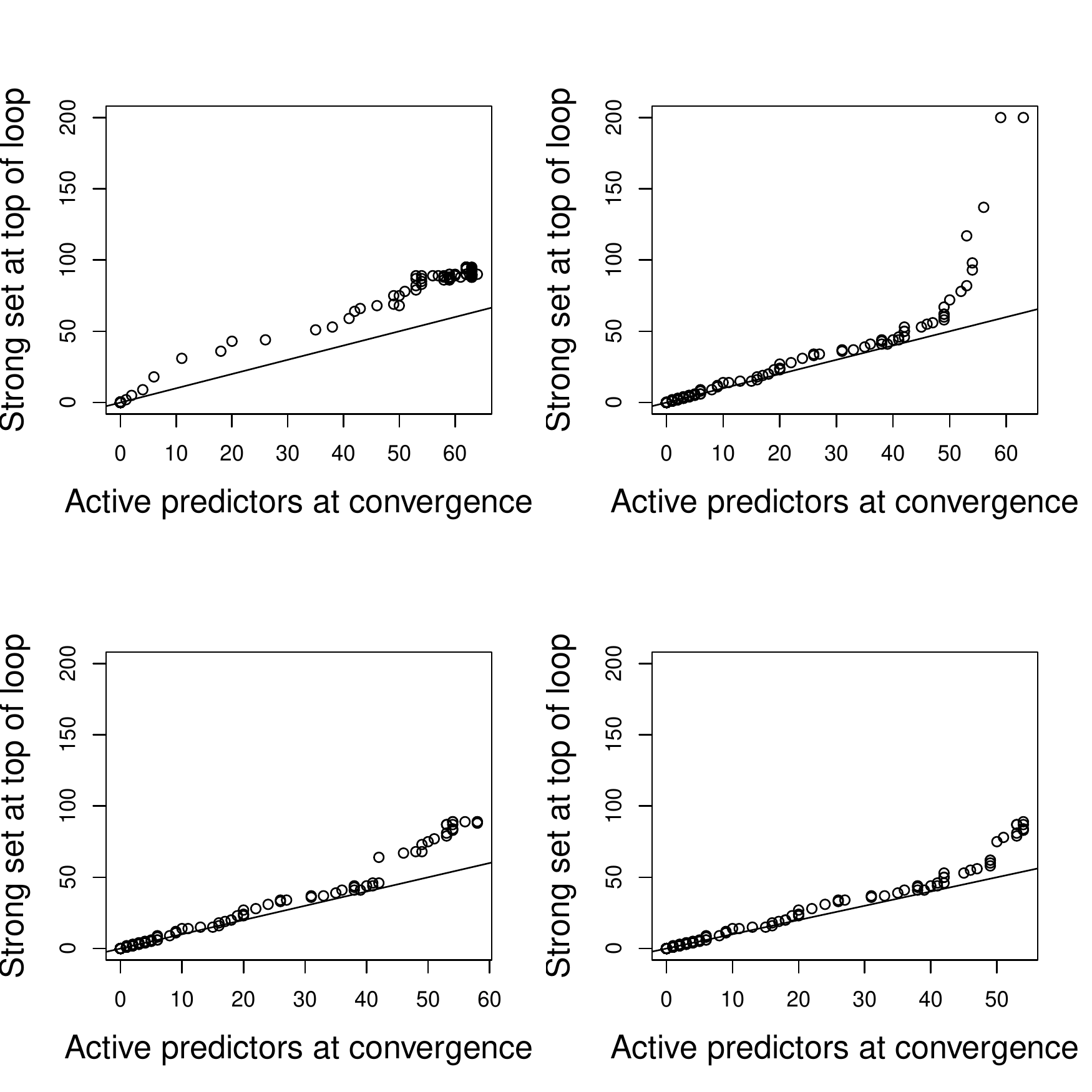}
    \caption{\emph{Screening performance of sequential strong rule for different grid choices. Top left panel: logarithmic jumps, top right: linear jumps, bottom left: 80 linear jumps followed by 20 logarithmic jumps, bottom right: 90 linear jumps followed by 10 logarithmic jumps. $p = 200$.}}
    \label{figScreenDiffGrid}
  \end{figure}

 \begin{table}[ht]
    \centering
    \begin{tabular}{c c c c c c}
      \hline\hline
      Jumps & $p = 100$ & $p = 200$ & $p = 500$ & $p = 1000$ & $p = 2000$ \\ [0.5ex]
      \hline
      Logarithmic & 0.50 & 0.77 & 2.19 & 3.93 & 7.65 \\
      Linear & 0.90 & 1.75 & 13.04 & 27.10 & 166.90 \\ 
      80 Linear; 20 Logarithmic & 0.46 & 0.70 & 2.00 & 3.65 & 7.05 \\
      90 Linear; 10 Logarithmic & 0.40 & 0.57 & 1.47 & 2.48 & 4.96 \\[1ex] 
      \hline
    \end{tabular}
    \caption{\emph{Speed in seconds for strong rules with different grid types}.}
    \label{tabDiffGrids}
  \end{table}
Notice the considerable speed up provided by the hybrid grids and the very poor performance of the linear grid at higher values of $p$. Figure~\ref{figScreenDiffGrid} demonstrates the screening performance of the sequential strong rule under each of the four grids. From the figure we see that the poor performance of the linear grid stems from its tendency to include nearly all of the predictors in the strong set at later iterations. Nearly all of the computation time is spent at these last one or two iterations.

Speed ups in the hybrid grids stem from at least three sources:
\begin{enumerate}
 \item Strong sets are smaller and fewer score and Hessian entries are computed at each iteration. This is demonstrated by the string of points in the third and fourth panels hugging the reference line more tightly than do those in the first panel.
 \item $\lambda$ values in the grid decrease more slowly under linear jumps, allowing the algorithm to spend more time where solutions are naturally sparse.
 \item $\lambda$ is not allowed to become as small as in the logarithmic case, so we never reach $\lambda$ values that would require a large number of non-zero parameters.
\end{enumerate}

No strong set violations were encountered for any of the grid choices over any of the simulations.

\section{Timings}
Having settled on the 90-10 linear-log grid, we ran further simulations in the same vein of Section 3, studying the performance of the algorithm over changes in the parameters.

\subsection{Fixed $K$}
Consider the case where the number of strata $K = 10$ is fixed. $B = 3$ simulation runs were made over a grid of parameters defined by:
\begin{itemize}
  \item $n = 10, 20, 40$.
  \item $m = 1, 5, \frac{n}{2}$.
  \item $p = 100, 200, 500, 1000, 2000$.
\end{itemize}

\begin{table}[ht]
    \centering
    \begin{tabular}{r r r r r r}
      \hline\hline
       & $p = 100$ & $p = 200$ & $p = 500$ & $p = 1000$ & $p = 2000$ \\ [0.5ex]
      \hline
      $n = 10; \text{ } m = 5$ & 0.33 & 0.70 & 1.25 & 2.49 & 4.65 \\
      $n = 20; m = 10$ & 0.83 & 1.78 & 3.97 & 8.38 & 27.71 \\ 
      $n = 40; m = 20$ & 3.87 & 9.94 & 41.35 & 180.28 & 433.76 \\[1ex] 
      \hline
      $n = 10; m = 1$ & 0.07 & 0.12 & 0.16 & 0.35 & 0.55 \\
      $n = 20; m = 1$ & 0.18 & 0.26 & 0.48 & 0.97 & 1.35 \\ 
      $n = 40; m = 1$ & 0.38 & 0.69 & 1.02 & 1.80 & 3.41 \\[1ex]
      \hline
      $n = 10; m = 5$ & 0.33 & 0.70 & 1.25 & 2.49 & 4.84 \\
      $n = 20; m = 5$ & 1.48 & 2.80 & 6.71 & 13.64 & 20.19 \\ 
      $n = 40; m = 5$ & 10.84 & 7.27 & 19.83 & 41.27 & 73.69 \\[1ex]
      \hline
    \end{tabular}
    \caption{\emph{Speed in seconds with fixed $K$}.}
    \label{tabFixedK}
  \end{table}

Table~\ref{tabFixedK} summarises the results. Evidently the $m = \frac{n}{2}$ case is most computationally intensive. This is as expected, since the normalising constant has the largest number of terms when $m$ has this relationship to $n$. Broad intuition about these numbers seems to be bourne out: running time seems to grow roughly linearly in $p$, linearly in $n$ for fixed $m$ and quadratically in $n$ if $m = \frac{n}{2}$. 

\subsection{Fixed sample size}
Since the total sample size $K \times n$ comprises of the number of strata multiplied by the number of observations within a stratum, we can run simulations wherein we keep the sample size fixed, but trade off between the two components. This provides a deeper understanding of how $K$ and $n$ affect the running time. 

Table~\ref{tabFixedSS} shows the results for two fixed sample sizes ($100$ and $200$) with rows showing how we trade off between $K$ and $n$. The number of cases within each stratum is taken to be $m = \lceil\frac{n}{2}\rceil$.

\begin{table}[ht] 
    \centering
    \begin{tabular}{r r r r r r}
      \hline\hline
      Sample size $= 100$ & $p = 100$ & $p = 200$ & $p = 500$ & $p = 1000$ & $p = 2000$ \\ [0.5ex]
      \hline
      $K = 50; \text{  }n = 2$ & 0.11 & 0.21 & 0.45 & 0.69 & 1.01 \\
      $K = 20; \text{  }n = 5$ & 0.19 & 0.39 & 0.87 & 1.50 & 2.86 \\ 
      $K = 10; n = 10$ & 0.37 & 0.64 & 1.55 & 2.08 & 3.84 \\
      $K = 5; n = 20$ & 0.29 & 0.59 & 1.14 & 2.00 & 3.74 \\
      $K = 2; n = 50$ & 0.61 & 1.18 & 2.32 & 4.64 & 9.60 \\   [1ex] 
      \hline
      Sample size $= 200$ & $p = 100$ & $p = 200$ & $p = 500$ & $p = 1000$ & $p = 2000$ \\ [0.5ex]
      \hline
      $K = 100; \text{ }n = 2$ & 0.36 & 0.67 & 1.45 & 2.96 & 6.21 \\
      $K = 50; \text{ }n = 4$ & 0.54 & 1.20 & 3.40 & 4.89 & 9.25 \\ 
      $K = 20; n = 10$ & 1.18 & 2.40 & 6.20 & 12.92 & 21.92 \\
      $K = 10; n = 20$ & 0.67 & 1.85 & 4.28 & 7.20 & 16.53 \\
      $K = 5; n = 40$ & 1.68 & 3.32 & 9.82 & 21.00 & 70.28 \\[1ex]
      \hline
    \end{tabular}
    \caption{\emph{Speed in seconds with fixed sample size}.}
    \label{tabFixedSS}
  \end{table}

Looking down the rows of the table, we see that running time seems to increase roughly linearly in $p$. The columns, however, tell a more interesting story. Notice how running times initially increase as the stratum sizes increase, but then seem to decrease once the number of strata becomes small enough (cancelling the within stratum effect). Finally, once strata become big enough, as in the last row of each segment of the table, the very large number of terms in the normalising constant requires a dramatic increase in computation. Running times are large here even though there are very few strata.

\subsection[Comparison to penalized package]{Comparison to \pkg{penalized} package}
\citet{Avalos2011} note the equivalence of the conditional logistic likelihood Equation~\ref{eq1.2} to the discrete partial likelihood of a Cox proportional hazards model with tied death times. In particular, for each stratum we associate cases with deaths and controls with censorings. All deaths occur at the same time (say time 1) and all censorings at some later time. The Cox discrete partial likelihood is exactly the same as the conditional probability contribution for a stratum in this case. Multiplying $K$ partial likelihoods together allows us to use software for the Cox likelihood to compute parameter estimates for the conditional likelihood. They suggest that the \proglang{R} package \pkg{penalized} of \citet{penalized} be used to fit the conditional logisitic model.

However, this package uses Breslow's approximation to deal with tied death times. Although a reasonable approximation for few ties, it quickly becomes poor if the number of ties grows relative to the total number of observations. So this method does not provide estimates gleaned from the exact likelihood. Furthermore, these estimates can differ quite dramatically from those gleaned from the exact likelihood even for moderate values of $m$. Opportunities for comparison between the \pkg{penalized} method and our method are thus limited, except when we have only 1 case in each stratum. Then the partial and conditional likelihoods are the same.

Table~\ref{tabPenal} shows the performance of the two methods (ours and one using \pkg{penalized}) for a simulation run as in Section 3. The number of strata is fixed at $K = 10$, each containing $m = 1$ case. Running times are quoted over a grid of different $n$-$p$  combinations.

\begin{table}[ht]
    \centering
    \begin{tabular}{c c c c c c}
      \hline\hline
      \pkg{clogitL1} & $p = 100$ & $p = 200$ & $p = 500$ & $p = 1000$ & $p = 2000$ \\ [0.5ex]
      \hline
      $n = 10$ & 0.07 & 0.10 & 0.20 & 0.39 & 0.77 \\
      $n = 20$ & 0.16 & 0.28 & 0.43 & 0.81 & 1.33 \\ 
      $n = 30$ & 0.23 & 0.37 & 0.78 & 1.40 & 2.41 \\
      $n = 40$ & 0.43 & 0.61 & 1.07 & 2.18 & 3.38 \\   [1ex] 
      \hline
      \pkg{penalized} & $p = 100$ & $p = 200$ & $p = 500$ & $p = 1000$ & $p = 2000$\\ [0.5ex]
      \hline
      $n = 10$ & 1.03 & 1.18 & 1.43 & 2.23 & 3.87 \\
      $n = 20$ & 1.19 & 1.47 & 1.93 & 3.13 & 5.35 \\ 
      $n = 30$ & 1.39 & 1.59 & 2.52 & 3.94 & 7.00 \\
      $n = 40$ & 1.83 & 1.99 & 2.78 & 5.22 & 8.90 \\   [1ex] 
      \hline
    \end{tabular}
    \caption{\emph{Speed in seconds for different algorithms: $K = 10$}.}
    \label{tabPenal}
  \end{table}
Our method seems to run more efficiently than does the \pkg{penalized} equivalent. This provides some comparison between the two algorithms for the (limited) case where both yield the same result.

\section{Comparison to unconditional model}
The conditional logistic regression model was initially promulgated to find estimates of $\beta$ in the logistic regression setup of Equation~\ref{eq1.1} where we have an intercept for each stratum. The conditioning argument eliminates nuisance intercept parameters from the objective, allowing the optimisation algorithm to focus attention only on estimating $\beta$, perhaps reducing bias and variance of these estimates. One could, of course, simply apply the unconditional (standard) logistic regression model, including dummy variables for each stratum to allow for estimation of the $\beta_{0k}$.

The question arises then about how these two methods compare, both in their variable selection performance and their ultimate predictive ability. This section seeks to address these via an appropriate simulation study.

\subsection{Simulation setup}
Datasets for each simulation run were generated as in the rest of the paper. We used the \textit{sample} function in \proglang{R}, which allows the selection of $m$ case indices from $\{1, 2, ..., n\}$ without replacement. It has a \textit{prob} argument which allows one to stipulate the weights with which this sample should be selected.

Regressors $X_{kij}$ were generated iid from a $N(0,1)$ distribution for $k = 1, 2,..., K$; $i = 1, 2, ..., n$ and $j = 1, 2,..., p$ for prespecified number of strata, $K$, observations per stratum, $n$, and regressors, $p$. Of these regressors, $q = \lfloor 0.1 \times p \rfloor$ had non-zero predictors with $\beta_j = \pm2$, the sign being selected uniformly at random. Intercepts $\beta_{0k}$ - one for each stratum - were generated iid from $N(0,2)$. Individual success probabilities were then obtained as in Equation~\ref{eq1.1} and fed to our sampling function to deliver $m$ case indices within each stratum.

For each of $B = 100$ simulated datasets, 4 models were fit, obtaining estimates of $\beta$ at a grid of 100 $\lambda$ values:

\begin{enumerate}
  \item Conditional logistic (using our software).
  \item Unconditional logistic without intercept.
  \item Unconditional logistic with a single intercept for all strata.
  \item Unconditional logistic with different intercept for each stratum.
\end{enumerate}
The latter three models were all fit using the \pkg{glmnet} package of \citet{glmnet}.

\subsection{Variable selection}
Although the conditional and unconditional models both seek to estimate $\beta$, they do so by maximising different criteria. Estimates will differ between the two types of model, with different numbers of non-zero parameters at similar levels of regularisation. 

For each $\lambda$ value, we can compute, for each method 1 - 4 above, a sensitivity (proportion of true non-zero $\beta$ correctly detected) and a specificity (proportion of true zero $\beta$ correctly left undetected). Ranging over the 100 $\lambda$ values for each method traces out a Receiver Operating Characteristic (ROC) curve. Since there are only a finite number of sensitivity values that can be achieved, an average ROC curve can be obtained by averaging the specificity values at each unique sensitivity value. These curves could provide some insight into the variable selection ability of each of the models. Data was generated in a way that clearly favours the conditional and multi-intercept unconditional models.

\begin{figure}[htb]
  \centering
  \includegraphics[height=100mm, width=100mm]{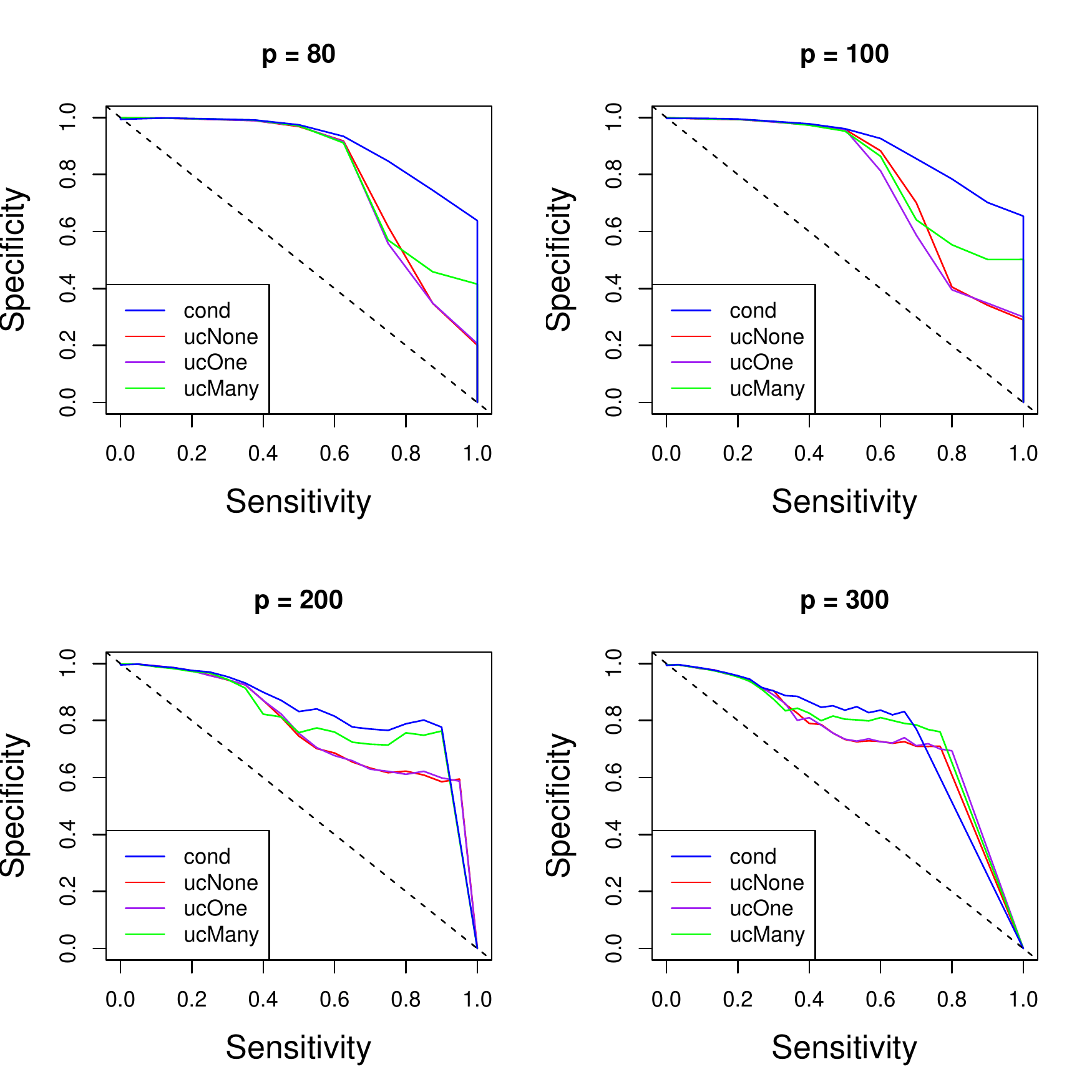}
  \caption {\emph{ROC curves for fixed n. Average ROC curves are plotted for conditional model (blue), unconditional model with no intercept (red), unconditional model with single intercept (purple) and unconditional model with intercept for every stratum (green). n = 2, m = 1, K = 75, p ranges over 80 (top left), 100 (top right), 200 (bottom left) and 300 (bottom right).}}
  \label{rocFixedn}
\end{figure}

\begin{figure}[htb]
  \centering 
  \includegraphics[height=100mm, width=100mm]{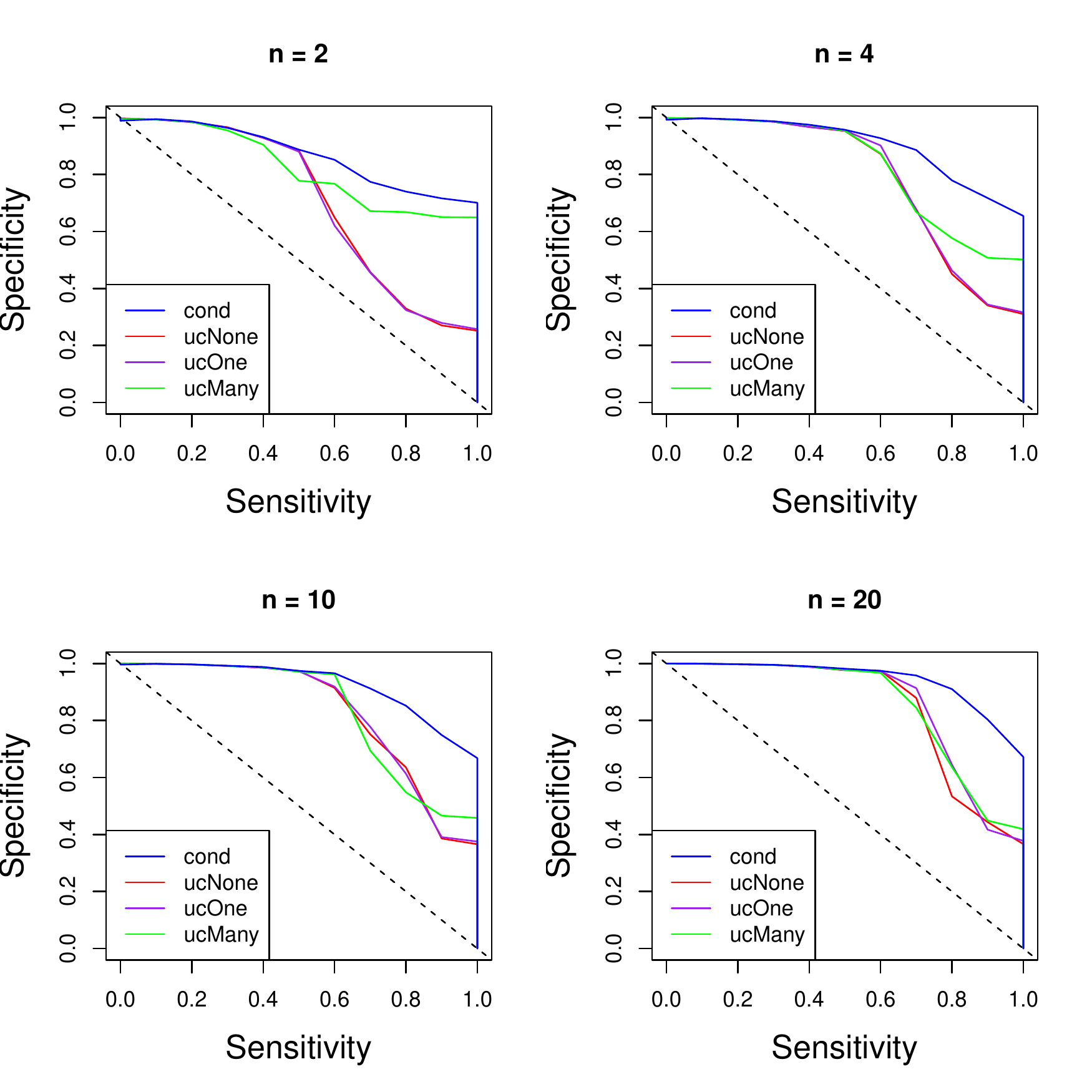}
  \caption{\emph{ROC curves for fixed p. Average ROC curves are plotted for conditional model (blue), unconditional model with no intercept (red), unconditional model with single intercept (purple) and unconditional model with intercept for every stratum (green). p = 100, n ranges over 2 (top left), 4 (top right), 10 (bottom left) and 20 (bottom right). Sample sizes are kept as close to 150 as possible, $m = \frac{n}{2}$}.}
  \label{rocFixedp}
\end{figure}

Figures~\ref{rocFixedn} and~\ref{rocFixedp} show the ROC curves obtained in our simulations. Sensitivity is on the horizontal axis; specificity on the vertical. The former figure fixes the number of observations $n$ within each stratum and varies the number of regressors $p$, while the latter fixes $p$ and allows $n$ to vary. Sample sizes are kept fixed at 150 and $m = \lfloor \frac{n}{2} \rfloor$. 

It seems as though the conditional and multi-intercept unconditional models do indeed offer better variable selection performance. Perhaps the conditional model, in turn, has an edge over the multi-intercept unconditional one. The difference between the conditional and unconditional models is greatest when strata are large and regressors are few and disappears quite quickly in $p$. Notice that the difference is quite significant for large $n$.

\subsection{Prediction}
One could ask how the superior variable selection performance is translated into predictive power. The trouble is that the conditional logistic model is not geared toward prediction. We make conditional arguments to eliminate intercepts as nuisance parameters, but these are crucial for predicting the class membership of new data points.

Even the unconditional model struggles here, because every new observation will come from a new stratum with a different intercept and we cannot learn these from the training data directly. At least this method provides estimates of the intercepts (either a single one or one for each stratum).

In order for the conditional model to provide predictions, we need to finesse the intercept problem. To this end, we propose two heuristically motivated methods. Both rely on computing a threshold $t_k$ for each stratum in the training set. All observations with linear predictor $(\hat{\beta}_\lambda^\top X_{ki})$ greater than the threshold are deemed cases; the rest controls. Thresholds are selected within each stratum to minimise the misclassification rate. A simple grid search over each of the observed linear predictors is sufficient to find the appropriate threshold.

Prediction of a new point proceeds by comparing its linear predictor to the average of the thresholds over all the training set strata (called the \textit{mean} method) or using each threshold to provide a prediction for the new test point and then taking a majority vote over all these predicitons (called the \textit{committee} method).

\begin {figure}[htb]
  \centering
  \includegraphics [height=100mm, width=100mm]{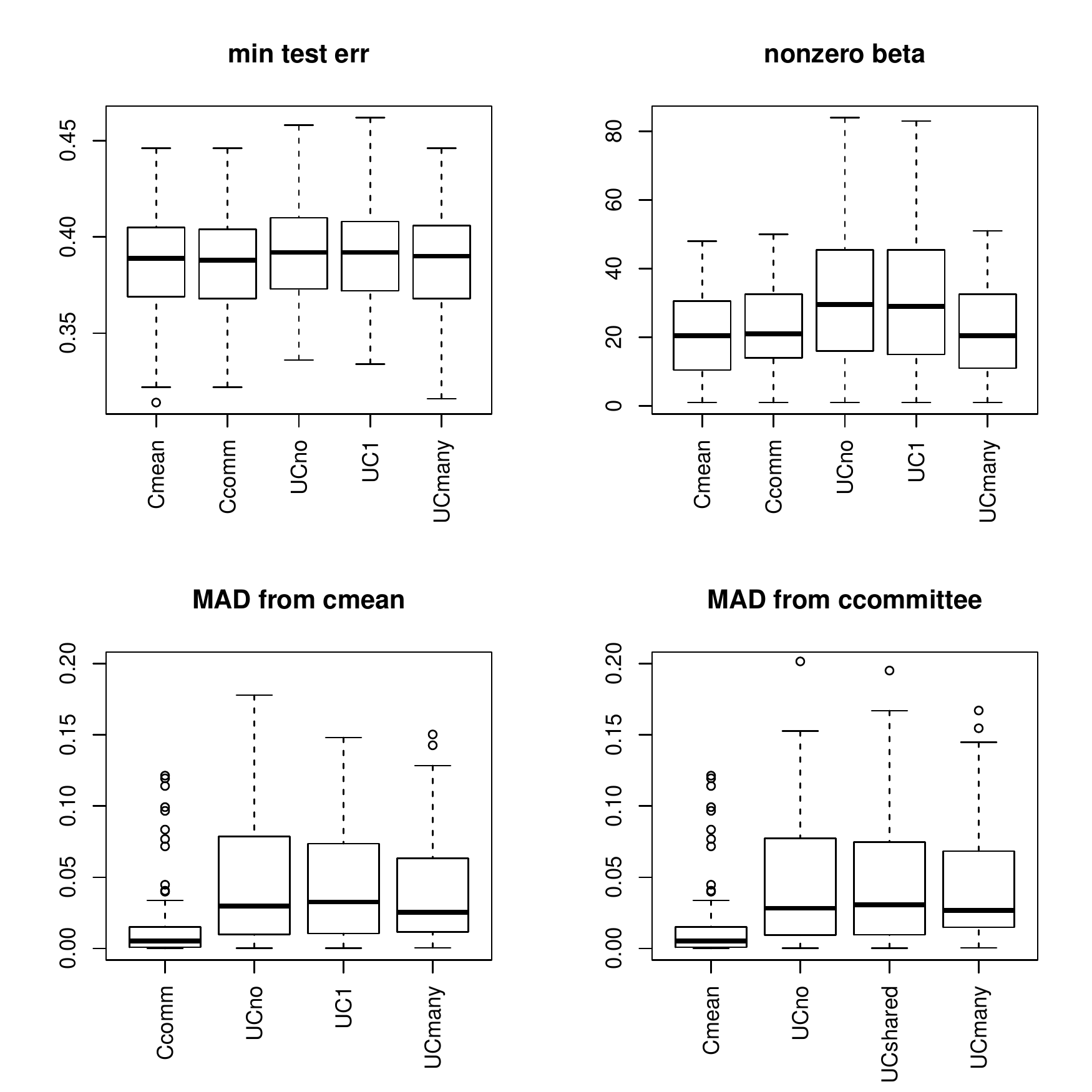}
  \caption{\emph{Boxplots summarising minimum test error parameter profiles for different methods. Top left: minimum test error. Top right: number of non-zero parameters at minimum test error. Bottom: mean absolute difference from conditional mean method (left) and committee methd (right) at minimum test error. p = 100, n = 2, m = 1.}}
  \label{predBox}
\end{figure}

We ran a simulation comparing the prediction performace of each of the methods. A training sample of size 150 was generated as above and a test set of size 750. The 4 models from the previous section were fit to the training set and used to predict the test set. For each simulated dataset, we could find $\lambda$ minimising test misclassification error for each method. This also gives us the $\beta$ profile of the test error minimising model. $B = 100$ simulations were run. Figure~\ref{predBox} shows boxplots summarising the results of these simulations. 

The top left panel shows the minimum test error for each of the methods. Notice how both conditional and the multi-intercept unconditional methods outperform (slightly) the other two unconditional methods. The top right panel shows the number of non-zero $\beta$ in the test error minimising profile for each method. Notice how the conditional methods seem to select slightly fewer variables.

The bottom two panels show the mean absolute difference from the \textit{mean} method $\beta$ profile (left) and the \textit{committee} method profile (right). It would seem that the two conditional methods produce very similar profiles, while conditional profiles differ somewhat from unconditional ones.

It should be noted that this is a difficult prediction problem:

\begin {itemize}
  \item Each stratum has its own intercept. Test points then have intercepts unrelated to those in the training set and it is difficult to obtain good estimates for them from the training set.
  \item Even if we knew the intercept of a test point exactly, we force each stratum of size $n$ to have $m = \lfloor \frac{n}{2} \rfloor$ cases. This means that even observations with small unconditional success probabilities (say below $0.5$) could be forced to be cases. This makes it even more difficult for these methods to predict, because they tend to predict cases when the unconditional success probability computed from the parameter estimates is above $0.5$.
\end{itemize}
Even though the prediction performance of the conditional methods and the multi-intercept unconditional method seems very similar, this may not speak to the quality of the $\beta$ estimates of each, but rather to both methods' inability to estimate intercepts well.

\subsection{Pathological solutions}
It is well known that the (unconstrained) unconditional logistic regression model produces pathological solutions when the data is perfectly separable. In particular, suppose that we have $N$ data points $(y_{i}, x_{i})$ of which $M$ are successes (cases). Also suppose we have only one regressor (so each $x_i \in \Re$). If $\max_{y_i = 1}x_i < \min_{y_i = 0}x_i$ or $\max_{y_i = 0}x_i < \min_{y_i = 1} x_i$, the maximum likelihood estimate of $\beta$ is driven to $\pm\infty$. There are $2 \times M! \times (N-M)!$ arrangements of the $x_i$ values that would lead to such a solution.

The problem is more severe for the conditional logistic model. Suppose that $N = \sum_{k=1}^K n_k$ and $M = \sum_{k=1}^K m_k$ so that we have $K$ strata with $n_k$ observations and $m_k$ cases apiece. Now the pathology occurs when we have separation (in the same direction) in each stratum, i.e., $\max_{y_{ki} = 0}x_{ki} < \min_{y_{ki} = 1} x_{ki}$ for all $k$ or $\max_{y_{ki} = 1}x_{ki} < \min_{y_{ki} = 0} x_{ki}$ for all $k$. The $2 \times M! \times (N-M)!$ arrangements above all satisfy these constraints and hence also lead to pathological solutions in the conditional model. There are, however, many more configurations that lead to separation (in the same direction) in each stratum.

Very large parameter estimates play havoc with optimisation algorithms, leading to instability and rounding errors. More care needs to be taken in the conditional case and the final solution will probably need to be regularised more severely than for the unconditional case.

\section{Cross validation}
Since intercepts are conditioned away in the conditional likelihood, we cannot use standard CV techniques to estimate prediction error. These standard techniques leave out some data, find parameter estimates and then predict the left out data. The average error over all the left out data serves as an estimate of the prediction error of the model. We choose the $\lambda$ that minimises this prediction error.

Here we follow the approach of \citet{HBH2006} and estimate the CV error of the $i^{th}$ left out fold to be:

\[
 \hat{CV}_i(\lambda) = l(\beta_{-i}(\lambda)) - l_{-i}(\beta_{-i}(\lambda))
\]
where $l_{-i}$ is the log-likelihood excluding part $i$ of the data and $\beta_{-i}(\lambda)$ is obtained by maximising $l_{-i} + \lambda||\beta||_1$.

Data should be split by leaving out entire strata at a time. These seem to be the most appropriate sampling units. Note that the above formula then has a very simple form: merely the sum over the left out strata of each stratum's log conditional likelihood contribution. We choose the $\lambda$ that maximises $\sum_i\hat{CV}_i(\lambda)$.

\begin{figure}[htb]
  \centering
  \includegraphics[height=100mm, width=100mm]{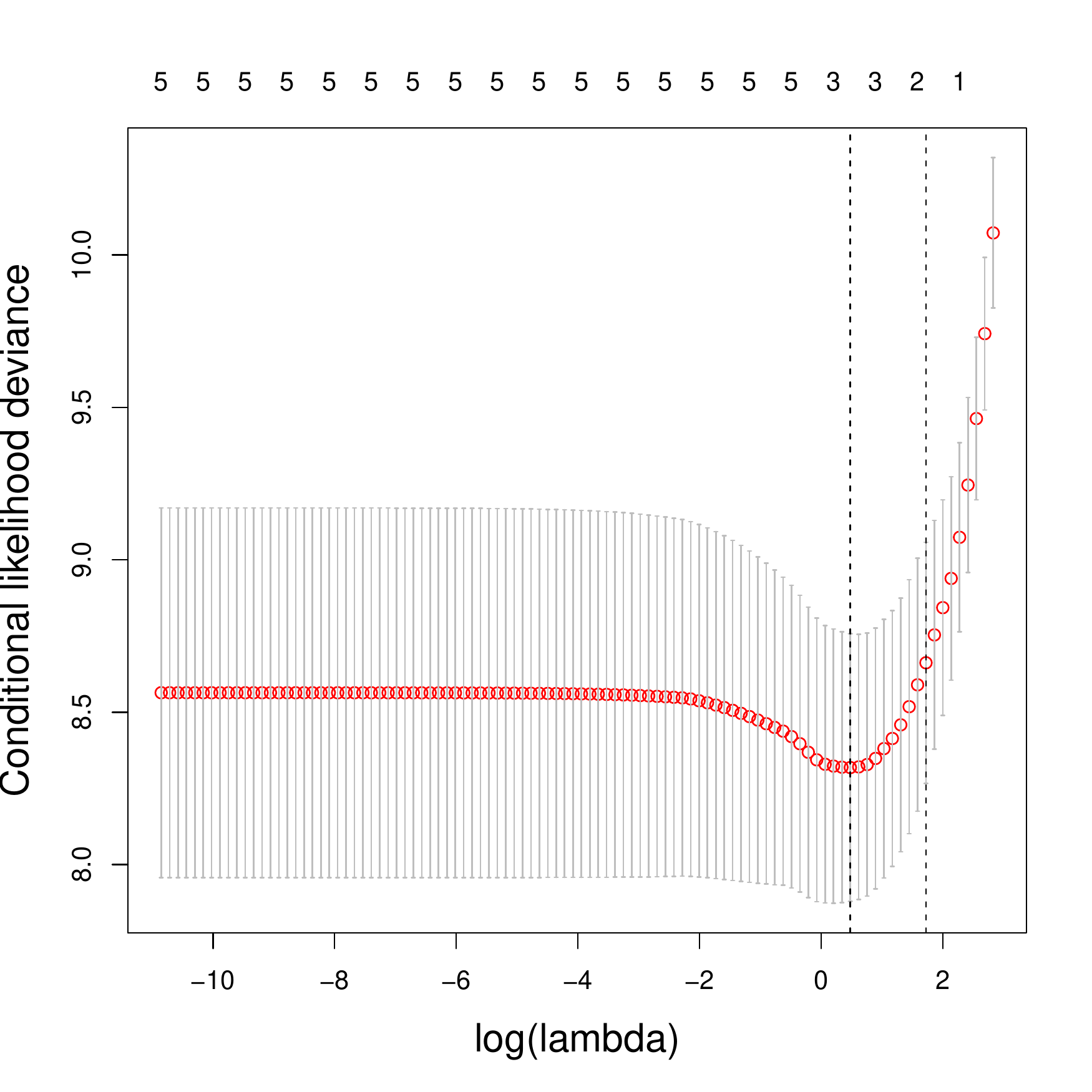}
  \caption{\emph{CV deviance curve (with standard error bands) for the endometrical cancer dataset. CV error seems to be minimised for a model with 3 predictors.}}
  \label{smallCV}
\end{figure}

As illustration, consider the small dataset of Section 2.1. This CV method was applied to it, using $10$ folds. Figure~\ref{smallCV} shows the CV curve (multiplied by $-1$ so we look for a minimum). Two vertical lines are drawn: the leftmost is at the minimising (log) $\lambda$, while the other is drawn at the smallest $\lambda$ with CV value one standard deviation away from the minimum CV error. The method seems to select $3$ predictors.

\section{Discussion}
We have applied the cyclic coordinate descent machinery to the fitting of conditional logistic regression models with lasso and elastic net penalties. Relatively efficient recursive algorithms were used to perform the computationally intensive task of computing normalising constants, gradients and Hessians for the likelihood within strata.

Sequential strong rules greatly improve efficiency by allowing us to compute fewer score and Hessian entires at every iteration and streamlining the computation of these dependent quantities.

A comparison was made between the conditional and unconditional models on a simulated dataset favouring the conditional model. The conditional model seems to perform admirably both in variable selection and predictive performance, outperforming the unconditional model in the former, while doing at least as well as the unconditional model at the latter.
\nocite{HP87}
\nocite{Epi}

\bibliography{jss1144}

\end{document}